\documentclass[11pt]{article}
\usepackage{moriond,epsfig}

\def\Journal#1#2#3#4{{#1} {\bf #2}, #3 (#4)}

\def\PRL{\em Phys. Rev. Lett.}
\def\PRD{{\em Phys. Rev.} D}

\def\be{\begin{equation}}
\def\ee{\end{equation}}
\def\bea{\begin{eqnarray}}
\def\eea{\end{eqnarray}}

\begin{document}
\vspace*{4cm}
\title{HEAVY FLAVOUR RESULTS FROM TEVATRON}

\author{ G. BORISSOV \\
(on behalf of the CDF and D\O\ collaborations)}

\address{Department of Physics, Lancaster University, \\
Lancaster LA1 4YB, England, UK}

\maketitle\abstracts{
The CDF and D\O\ experiments finalize the analysis of their full statistics
collected in the $p \bar p$ collisions at a center-of-mass energy of $\sqrt{s} = 1.96$ TeV
at the Fermilab Tevatron collider. This paper presents several new results on the properties
of hadrons containing heavy $b$- and $c$-quarks obtained by both collaborations.
These results include the search for the rare decays $B^0, B^0_s \to \mu^+ \mu^-$ (CDF),
the study of CP asymmetry in $B_s \to J\psi \phi$ decay (CDF, D\O),
the measurement of the like-sign dimuon charge asymmetry (D\O),
the measurement of CP asymmetry in $D^0 \to K^+K^-$ and $D^0 \to \pi^+\pi^-$ decays (CDF),
and the new measurement of the $B_s \to D_s^{(*)+} D_s^{(*)-}$ branching fraction (CDF).
Both experiments still expect to produce more results on the properties of heavy flavours.
}

For past 10 years the Fermilab Tevatron collider has pioneered and established the role of hadron
colliders for flavour physics. It became the main source of results on $B_s^0$, $B_c$ mesons,
and $B$ baryons. Many crucial measurements, like the mass difference in the $B_s^0$ system,
were obtained here. Currently the Tevatron experiments finalize their study and publish
the results with the full statistics up to 10 fb$^{-1}$.

In this paper I review
\begin{itemize}
\item search for the rare decays $B^0, B_s^0 \to \mu^+ \mu^-$ obtained by the CDF collaboration;
\item study of CP asymmetry in $B_s^0 \to J/\psi \phi$ decay reported by the CDF and D\O\ collaborations;
\item updated measurement of the like-sign dimuon charge asymmetry (D0 collaboration);
\item new measurement of the difference of CP asymmetry in
$D^0 \to K^+K^-$ and $D^0 \to \pi^+\pi^-$ decays (CDF collaboration);
\item measurement of the branching fraction of $B_s^0 \to Ds^{(*)+} Ds^{(*)-}$ decay (CDF collaboration).
\end{itemize}

The standard model (SM) predicts a very low value for the branching fractions of both
$B^0 \to \mu^+ \mu^-$ and $B_s^0 \to \mu^+ \mu^-$ decays. The most recent SM prediction
for these fractions is \cite{buras}
\begin{eqnarray}
Br(B_s^0 \to \mu^+ \mu^-) & = & (3.2 \pm 0.2) \times 10^{-9}, \nonumber \\
Br(B^0   \to \mu^+ \mu^-) & = & (1.0 \pm 0.1) \times 10^{-10}.
\end{eqnarray}
The contribution of new physics beyond the SM can significantly modify these values, therefore these rare
decays can provide important constraints on various new physics models.

\begin{figure}
\begin{center}
\epsfig{figure=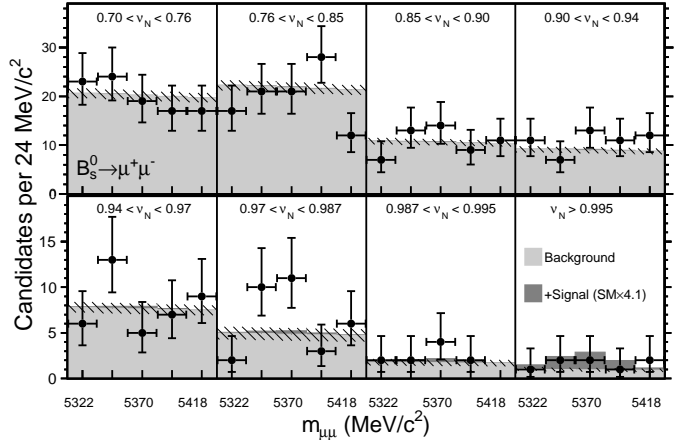,height=6cm}
\end{center}
\caption{
For the $B_s$ mass region, the observed number of events (points) is compared to the
total expected background (light grey) and its uncertainty (hatched) for different values of $\nu_N$.
The hashed area represents the systematic uncertainty on the mean expected background while the error
bars on the points represent the associated poisson uncertainty.
Also shown is the expected contribution from $B_s^0 \to \mu^+ \mu^-$ events (dark gray) using a
branching fraction that corresponds to the central value from the fit to the data,
which is 4.1 times the expected SM value.
\label{fig1}}
\end{figure}

The CDF collaboration presented in summer 2011 the analysis \cite{cdf-mumu-1}
with 7 fb$^{-1}$ featuring an accumulation of signal-like events in the $B_s^0$ mass region
with $\sim 2.5 \sigma$ deviation from the background-only hypothesis.
The new CDF analysis presented here includes the full Run2 statistics corresponding
to 9.6 fb$^{-1}$. Given the increased interest to the previous result,
the analysis of the remaining statistics is kept the same. The separation between the signal and
background in this analysis is achieved using the neural network. Figure \ref{fig1} shows the observed
and expected number of events in the $B_s^0 \to \mu^+ \mu^-$ search
for the different values of the neural network output variable $\nu_N$.
There is an excess of the signal-like events for $\nu_N > 0.97$,
while the agreement between the observed and expected number of events is very good for
the background-dominated region $\nu_N < 0.97$. The $p$-value of the SM signal plus background hypothesis for
$\nu_n > 0.97$ is 7\%. The excess of events in the $0.97 < \nu_N < 0.987$ bin is not increased with the
addition of the new statistics and is consistent with the statistical fluctuation. The $p$-value
of the SM signal plus background hypothesis for two largest $\nu_N$ bins is 22.4\%, while the $p$-value of
background only hypothesis is 2.1\%. Thus, while still not conclusive, the experiment becomes
sensitive to the SM contribution of $B_s^0 \to \mu^+ \mu^-$ decay and shows a good agreement with
the SM expectation.

The results obtained by the CDF collaboration with 9.6 fb$^{-1}$ are:
\begin{eqnarray}
Br(B_s^0 \to \mu^+ \mu^-) & = & (1.3^{+0.9}_{-0.7}) \times 10^{-8}, \nonumber \\
Br(B^0   \to \mu^+ \mu^-) & < & 4.6 \times 10^{-9}~ (3.8 \times 10^{-9})~ \mbox{at 95\% (90\%) C.L.}
\end{eqnarray}
The CDF collaboration also reports the first double sided limit on $Br(B_s^0 \to \mu^+ \mu^-)$:
\begin{eqnarray}
0.8 \times 10^{-9}  < & Br(B_s^0 \to \mu^+ \mu^-) & <  3.4 \times 10^{-8} ~ \mbox{at 95\% C.L.}, \nonumber \\
2.2 \times 10^{-9}  < & Br(B_s^0 \to \mu^+ \mu^-) & <  3.0 \times 10^{-8} ~ \mbox{at 90\% C.L.}
\end{eqnarray}
These results are consistent with other searches of these rare decays.

An important part of the research of heavy flavours at hadron colliders is devoted to the
measurement of the $CP$ asymmetry.
Among other reasons, the interest to this phenomenon is explained by the fact that
the magnitude of the $CP$ asymmetry included in the SM is not sufficient to describe the observed abundance
of matter in our universe \cite{huet}, which implies that some additional sources
of $CP$ asymmetry should exist.
They could reveal themselves by the deviation of the observed $CP$ asymmetry from the SM prediction.

One of the most promising channels to search for the new sources of $CP$ asymmetry is the
decay $B_s^0 \to J/\psi \phi$. The $CP$ asymmetry in this decay is described by the phase $\phi^{J/\psi \phi}$.
Within the SM, this phase is related with the angle $\beta_s$ of the $(bs)$ unitarity triangle
and is predicted to be very small \cite{lenz}:
\begin{equation}
\phi^{J/\psi \phi}(SM) = -2 \beta_s = -0.036 \pm 0.002.
\end{equation}
This phase can be significantly modified by the new physics contribution and this deviation from the SM
can be detected experimentally.

\begin{figure}
\begin{center}
\epsfig{figure=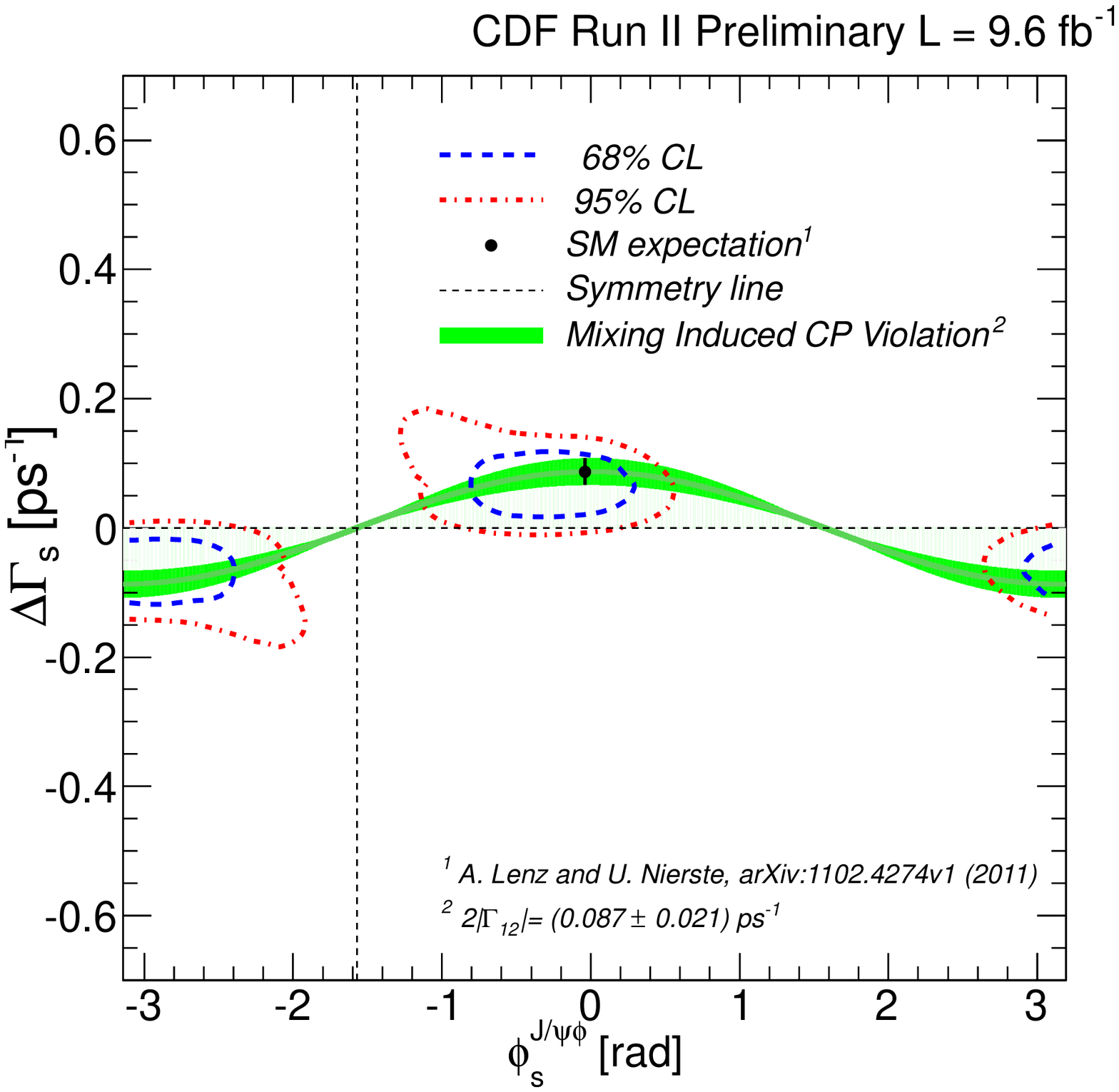,height=7cm}
\epsfig{figure=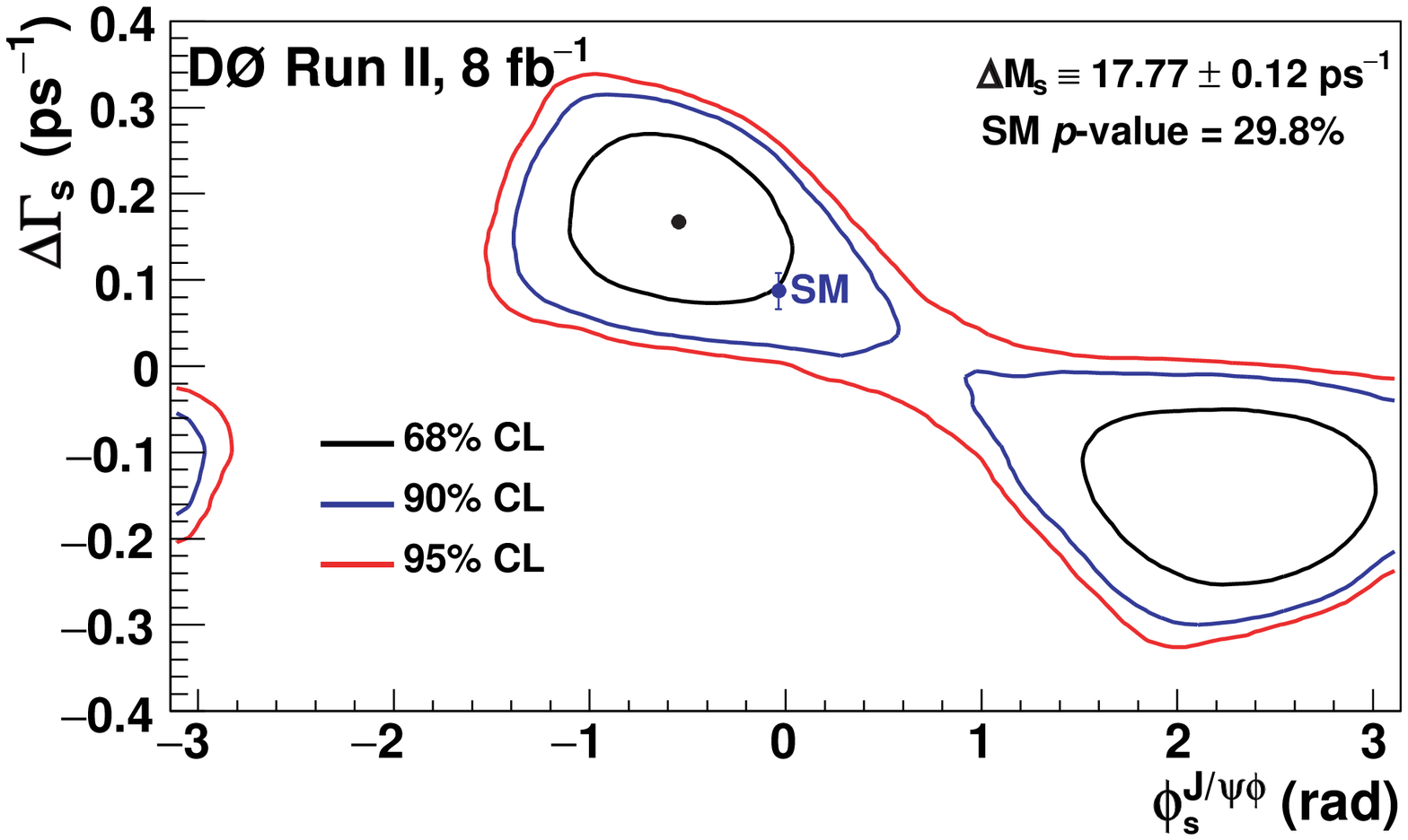,height=5cm}
\end{center}
\caption{
Confidence regions in $\phi^{J/\psi \phi} - \Delta \Gamma_s$ plane.
Left plot: the measurement of CFT collaboration.
The solid (blue) and dot-dashed (red) contours show the 68\% and
95\% confidence regions, respectively. The shaded (green) band is the theoretical
prediction of mixing-induced $CP$ asymmetry.
Right plot: the measurement of D\O\ collaboration. Two-dimensional 68\%, 90\% and 95\%
C.L. contours including systematic uncertainties are shown.
The standard model expectation is indicated as a point with an error.
\label{fig2}}
\end{figure}

Both CDF and D\O\ experiments report the new study of $B_s^0 \to J/\psi \phi$ decay with the full statistics.
The CDF collaboration reconstructs about 11000 such decays using the integrated
luminosity 9.6 fb$^{-1}$. The new analysis \cite{cdf-jpsi} is similar to
the previous measurement with a part of the statistics \cite{cdf-jpsi-1}. The result of this analysis
is presented in Fig. \ref{fig2} (left plot) as the confidence regions in $\phi^{J/\psi \phi} - \Delta \Gamma_s$ plane.
It can be seen that the obtained confidence region is consistent with the SM prediction within 1$\sigma$.
The obtained confidence regions for the quantity $\beta_s^{J/\psi \phi} \equiv -\phi^{J/\psi \phi}/2$ is
\begin{eqnarray}
\beta_s^{J/\psi \phi} & \in & [-\pi/2, -1.51] \cup [-0.06, 0.30] \cup [1.26, \pi/2] ~ \mbox{at 68\% C.L.} \nonumber \\
\beta_s^{J/\psi \phi} & \in & [-\pi/2, -1.36] \cup [-0.21, 0.53] \cup [1.04, \pi/2] ~ \mbox{at 95\% C.L.}
\end{eqnarray}

A similar analysis of $B_s^0 \to J/\psi \phi$ decay by the D\O\ collaboration \cite{d0-jpsi}
is based on 6500 signal events
collected using the integrated luminosity 8 fb$^{-1}$. The result of this analysis
is shown in Fig. \ref{fig2} (right plot). The obtained confidence region is consistent
with the SM prediction, and the $p$-value for the SM point is 29.8\%.
The following values are obtained in this analysis:
\begin{eqnarray}
\tau_s & = & 1.443_{-0.035}^{+0.038}~\mbox{ps}, \nonumber \\
\Delta \Gamma_s & = & 0.163^{+0.065}_{-0.064} ~\mbox{ps}^{-1}, \nonumber \\
\phi^{J/\psi \phi} & = & -0.55^{+0.38}_{-0.36}.
\end{eqnarray}

Another quantity sensitive to the new sources of the $CP$ asymmetry
is the like-sign dimuon charge asymmetry, which is defined as
\begin{equation}
A_{\rm sl}^b \equiv \frac{N_b^{++}-N_b^{--}}{N_b^{++}+N_b^{--}}.
\end{equation}
Here $N_b^{++}$ and $N_b^{--}$ represent the number of events containing two $b$ hadrons
decaying semileptonically and producing two positive or two negative muons, respectively.
The standard model predicts a very small value compared to the current experimental sensitivity,
therefore, the non-zero value of the asymmetry $A_{\rm sl}^b$ signals the presence of the $CP$ violation
in mixing in the semileptonic decays of neutral $B$ mesons. 
Recently the D\O\ collaboration released the new measurement of this quantity \cite{d0-asl}
using the integrated luminosity 9 fb$^{-1}$. The obtained value of $A_{\rm sl}^b$ deviates
from the SM prediction by 3.9 $\sigma$:
\begin{equation}
A_{\rm sl}^b = (-0.787 \pm 0.172 \pm 0.093)\%
\label{aslb}
\end{equation}

\begin{figure}
\begin{center}
\epsfig{figure=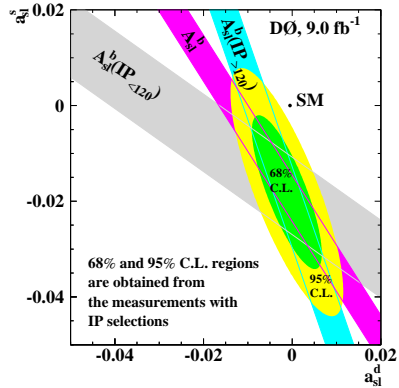,height=5.5cm}
\end{center}
\caption{
Measurement of $A_{\rm sl}^b$ with different muon IP selections in the $(a_{\rm sl}^d, a_{\rm sl}^s)$ plane.
The bands represent the $\pm 1$ standard deviation uncertainties on each individual measurement.
The ellipses represent 68\% and 95\% two-dimensional CL regions, respectively, of of $a_{\rm sl}^d$
and $a_{\rm sl}^s$ values obtained from the measurements with IP selections.
\label{fig3}}
\end{figure}

The asymmetry $A_{\rm sl}^b$ contains the contribution from the semileptonic charge asymmetries
$a_{\rm sl}^d$ and $a_{\rm sl}^s$ of $B^0$ and $B_s^0$ mesons, respectively. The analysis
of the dependence of $A_{\rm sl}^b$ on the muon impact parameter (IP) allows to obtain the separate
values of $a_{\rm sl}^d$ and $a_{\rm sl}^s$:
\begin{eqnarray}
a_{\rm sl}^d & = & (-0.12 \pm 0.52)\%, \nonumber \\
a_{\rm sl}^s & = & (-1.81 \pm 1.06)\%.
\end{eqnarray}
The precision of these quantities is comparable with the available world average measurements.
Figure \ref{fig3} presents the results of the IP study in the $(a_{\rm sl}^d, a_{\rm sl}^s)$ plane
together with the result (\ref{aslb}) of the $A_{\rm sl}^b$ measurement. The ellipses represent
the 68\% and 95\% two-dimensional confidence level (CL) regions, respectively, of $a_{\rm sl}^d$
and $a_{\rm sl}^s$ values obtained from the IP study. The obtained values of $a_{\rm sl}^d$ and $a_{\rm sl}^s$
are in a good agreement with the independent measurement of $a_{\rm sl}^s$ by the D\O\ collaboration \cite{d0-asls},
and the world-average value of $a_{\rm sl}^d$ reported by the HFAG \cite{hfag}. The discrepancy between the
measured value of $A_{\rm sl}^b$ and the SM prediction requires an independent confirmation.

One more promising channel to search for the new sources of $CP$ asymmetry is the single 
Cabibbo-suppressed decays $D^0 \to K^+K^-$ and $D^0 \to \pi^+ \pi^-$.
Although the exact theoretical prediction of $CP$ asymmetry in these decays
is difficult to obtain due to the non-perturbative contributions, the $CP$ asymmetry at $\mathcal{O}(1\%)$
level could signal the contribution of new physics.

The CDF collaboration previously measured \cite{cdf-d0} the separate values of asymmetries $A_{CP}(K^+K^-)$
and $A_{CP}(\pi^+\pi^-)$ with 6 fb$^{-1}$:
\begin{eqnarray}
A_{CP}(\pi^+\pi^-) & = & (+0.22 \pm 0.24 \pm 0.11) \%, \nonumber \\
A_{CP}(K^+K^-)     & = & (-0.24 \pm 0.22 \pm 0.09) \%.
\end{eqnarray}
The new analysis \cite{cdf-d0-2} uses the full data set corresponding to the luminosity 9.6 fb$^{-1}$
and is optimized for the measurement of $\Delta A_{CP} \equiv A_{CP}(K^+K^-) - A_{CP}(\pi^+\pi^-)$. It is
motivated by the recent result reported by the LHCb experiment \cite{lhcb-d0}:
\begin{equation}
\Delta A_{CP}\mbox{(LHCb)} = (-0.82 \pm 0.21 \pm 0.11) \%.
\label{a-lhcb}
\end{equation}

\begin{figure}
\begin{center}
\epsfig{figure=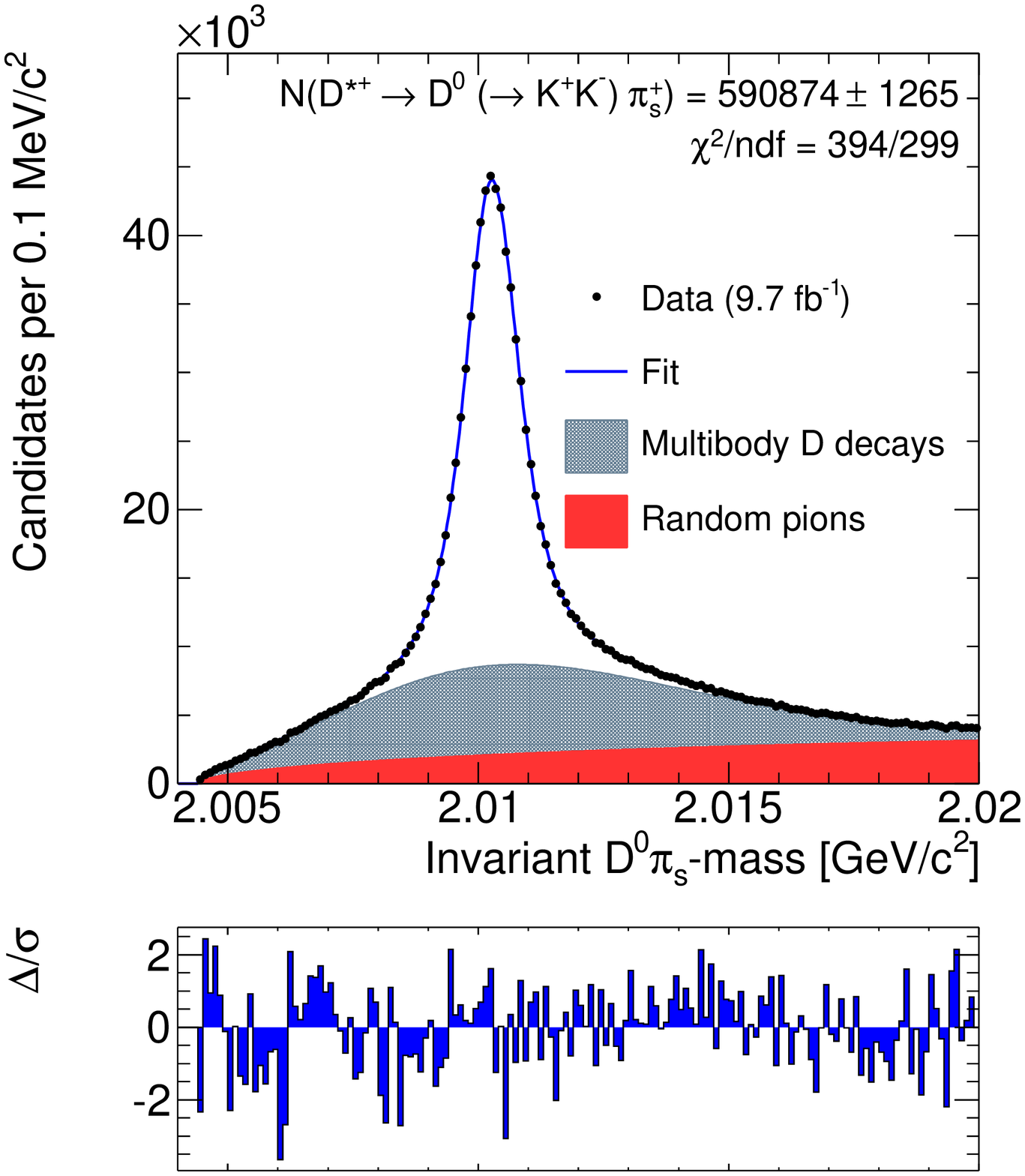,height=6cm}
\epsfig{figure=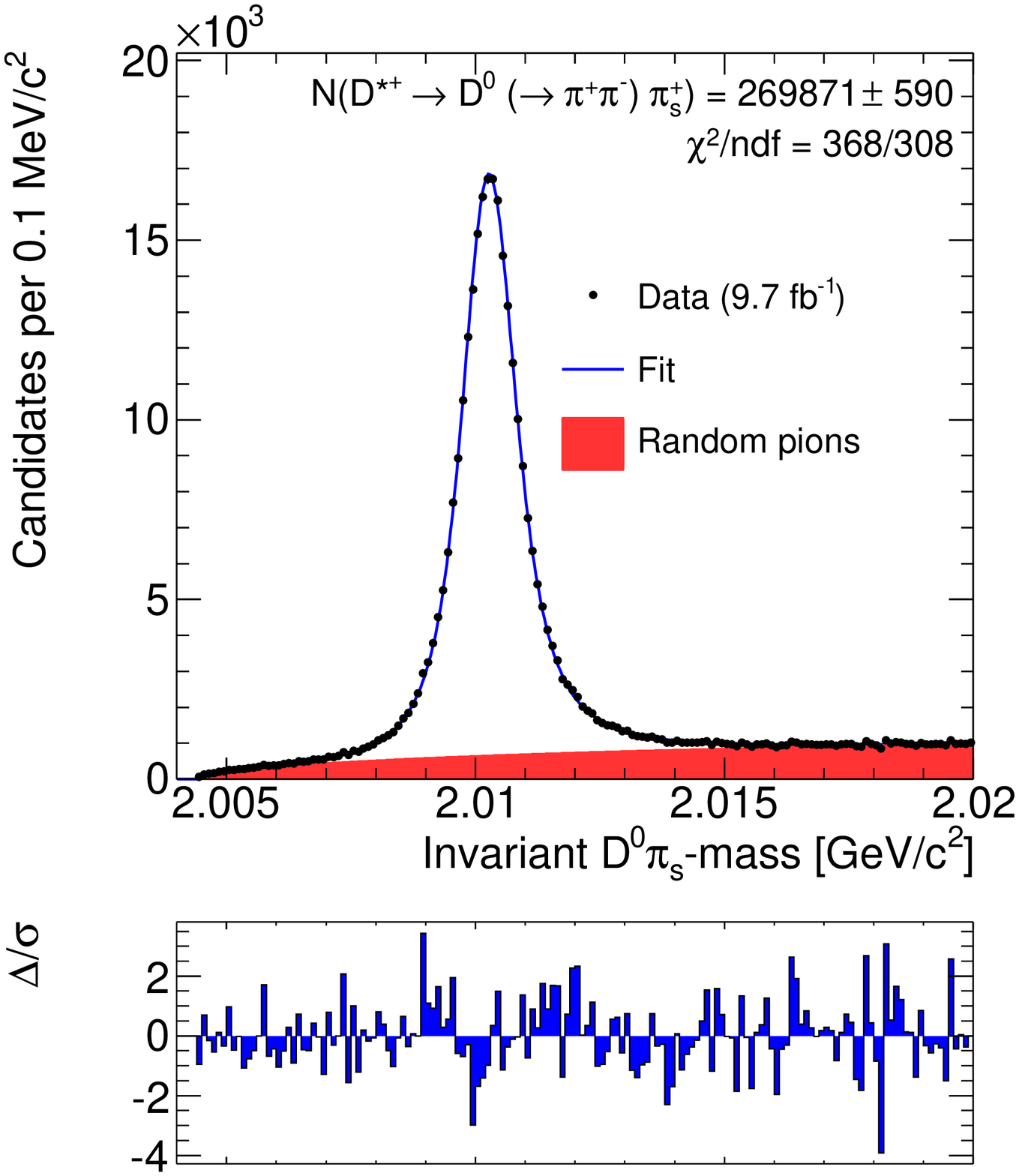,height=6cm}
\end{center}
\caption{
Projections of the combined fit on data for tagged $D^0 \to K^+ K^-$ (left)
and $D^0 \to \pi^+ \pi^-$ (right) decays.
\label{fig4}}
\end{figure}

Many systematic uncertainties cancel in the difference of asymmetries
and therefore the selection cuts in the new CDF analysis are loosened to increase the statistics.
In total 550K $D^0 \to \pi^+\pi^-$ decays and 1.21M $D^0 \to K^+K^-$ decays are selected.
Both decays are reconstructed in the $D^{*\pm} \to \stackrel{(-)}{D^0} \pi^\pm$ decay.
Figure \ref{fig4} shows the mass distributions of the reconstructed
$D^0 \to K^+K^-$ and $D^0 \to \pi^+ \pi^-$ decays. It can be seen that
the quality of the description of the data is excellent. Using the collected statistics,
the CDF collaboration obtains
\begin{equation}
\Delta A_{CP}\mbox{(CDF)} = (-0.62 \pm 0.21 \pm 0.10) \%,
\end{equation}
which corresponds to 2.7 $\sigma$ deviation from zero.
This result is consistent with the LHCb measurement (\ref{a-lhcb}). The combination of the CDF and LHCb results
gives $\sim 3.8 \sigma$ deviation of $\Delta A_{CP}$ from zero.

\begin{figure}
\begin{center}
\epsfig{figure=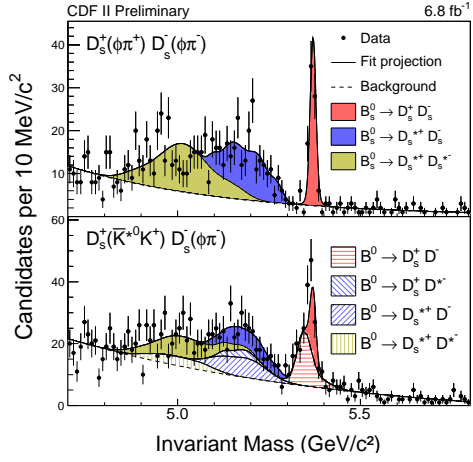,height=6cm}
\end{center}
\caption{
Invariant mass distribution of $B_s^0 \to D_s^{+}(\phi \pi^+) D_s^{-} (\phi \pi^-)$
and $B_s^0 \to D_s^{+}(K^{*0} K^+) D_s^{-} (\phi \pi^-)$.
\label{fig5}}
\end{figure}

The CDF collaboration reports \cite{cdf-dsds} one more interesting result on the properties of $B^0_s$
meson, namely the measurement of the branching fraction
$Br(B_s^0 \to D_s^{(*)+} D_s^{(*)-})$. It is obtained using the semi-exclusive decay modes
\begin{eqnarray}
B_s^0 & \to & D_s^{+} D_s^{-}, \nonumber \\
B_s^0 & \to & D_s^{*+} D_s^{-} + D_s^{+} D_s^{*-}, \nonumber \\
B_s^0 & \to & D_s^{*+} D_s^{*-},
\end{eqnarray}
with $D_s \to \phi \pi$ or $D_s \to K^* K$.
The resulting invariant mass distribution is presented in Fig.~\ref{fig5}.
In total 750 signal events in these decay modes are reconstructed.
Using this statistics, the following result is obtained
\begin{eqnarray}
Br(B_s^0 \to D_s^{+} D_s^{-}) & = & (0.49 \pm 0,06 \pm 0.05 \pm 0.08)\%, \nonumber \\
Br(B_s^0 \to D_s^{*+} D_s^{-} + D_s^{+} D_s^{*-}) & = & (1.13 \pm 0.12 \pm 0.09 \pm 0.19)\%, \nonumber \\
Br(B_s^0 \to D_s^{*+} D_s^{*-}) & = & (1.75 \pm 0.19 \pm 0.17 \pm 0.29)\%.
\end{eqnarray}
The total branching fraction of these decay modes is found to be
\begin{equation}
Br(B_s^0 \to D_s^{(*)+} D_s^{(*)-}) =  (3.38 \pm 0.25 \pm 0.30 \pm 0.56)\%.
\end{equation}

In conclusion, the experiments at the Tevatron finalize the analysis of their full statistics. This
paper presents the new results obtained in the search for the rare decays $B^0, B^0_s \to \mu^+ \mu^-$ (CDF),
the study of CP asymmetry in $B_s \to J\psi \phi$ decay (CDF, D\O),
the measurement of the like-sign dimuon charge asymmetry (D\O),
the measurement of CP asymmetry in $D^0 \to K^+K^-$ and $D^0 \to \pi^+\pi^-$ decays (CDF),
and the new measurement of the $B_s \to D_s^{(*)+} D_s^{(*)-}$ branching fraction (CDF).
Many more exciting results from the D\O\ and CDF experiments with the full statistics can be expected
soon.

\end{document}